\documentclass{iaus}
\usepackage{graphicx}

\title[Parameters for SMC clusters from CMD modelling] 
{Self-consistent physical parameters for MC clusters 
from CMD modelling: application to SMC clusters
observed with the SOAR telescope}

\author[Bruno Dias, Leandro Kerber, Bas\'{\i}lio Santiago {\it et al.}]   
{Bruno M. S. Dias$^1$, Leandro O. Kerber$^{1,2}$, Bas\'{\i}lio X. Santiago$^3$,
Beatriz Barbuy$^1$
 \and Eduardo Balbinot$^3$}

\affiliation{
$^1$IAG, Universidade de S\~ao Paulo, Rua do Mat\~ao, 1226, 
S\~ao Paulo, 05508-900, Brazil \\ 
email: {\tt (bdias,kerber,barbuy)@astro.iag.usp.br} \\
[\affilskip]
$^2$INAF - Osservatorio Astronomico di Padova, vic. Osservatorio, 5, I -
35122, Padova, Italy\\
email: {\tt leandro.kerber@oapd.inaf.it} \\
[\affilskip]
$^3$IF, Universidade Federal do Rio Grande do Sul, 
Av. Bento Gon\c calves, 9500, Porto Alegre, 91501-970, Brazil \\
email: {\tt (basilio.santiago,eduardo.balbinot)@ufrgs.br}
}

\pubyear{2008}
\volume{256}  
\pagerange{001--004}
\jname{The Magellanic System: Stars, Gas, and Galaxies}
\editors{Jacco Th. van Loon \& Joana M. Oliveira, eds.}
\begin{document}

\maketitle

\begin{abstract}
The Magellanic Clouds (MCs) present a rich system of stellar clusters 
that can be used to probe the dynamical and chemical evolution of these 
neighboring and interacting irregular galaxies. In particular, these 
stellar clusters (SCs) present combinations of age and metallicity that are 
not found for this class of objects in the Milky Way, being therefore 
very useful templates to test and to calibrate integrated light simple 
stellar population (SSP) models applied to unresolved distance galaxies. 
On its turn, the age and metallicity for a cluster can be determined 
spatially resolving its stars, by means of analysis of its 
colour-magnitude diagrams (CMDs). In this work we present our method 
to determine self-consistent physical parameters (age, metallicity, 
distance modulus and reddening) for a stellar cluster, from CMDs modelling 
of relatively unstudied SCs in the Small Magellanic Cloud (SMC) 
imaged in the BVI filters with the 4.1 m SOAR telescope.
Our preliminary results confirm 
our expectations that come from a previous integrated spectra and colour 
analysis: at least one of them (Lindsay~2) is an intermediate-age 
stellar cluster with $\sim$~2.6 Gyr and [Fe/H]~$\sim$~-1.3, 
being therefore a new interesting witness regarding the reactivation 
of the star formation in the MCs in the last 4 Gyr. 

\end{abstract}

\firstsection 

\section{Introduction}

SCs are useful tools to study the complex stellar 
content observed in nearby galaxies, as they may be modelled 
as SSP of a fixed age and metallicity.
In particular, the MC SCs form 
a rich system ($>$~3700 objects) (Bica et al. 2008), 
therefore they can be used to test and to calibrate SSP models
for combinations of age and metallicity that are not found in the
Milky Way (Santos Jr. \& Piatti 2004). Such information can be used to probe the
dynamical and chemical evolution of these neighboring and interacting 
dwarf irregular galaxies.

The age distribution based on clusters is likely distinct from the 
star formation history (SFH) as inferred from field stars (Holtzman 
et al. 1999). More recently, Rafelski \& Zaritsky (2005) 
analysed a sample of 195 clusters 
and shows that the populations of field stars are
similar to the populations of SMC stars.
The large period of quiescent star formation in the MCs 
between $\sim$~4 and 10~Gyr (Harris \& Zaritsky 2001, 2004) seems
to be imprinted in the low number of populous SCs 
with these ages (Rich et al. 2000, 2001). 
Almost all of them are in the SMC as well 
(Mighell et al. 1998; Piatti et al. 2005, 2007).

In spite of that the SMC cluster system is comparatively much less 
studied than the LMC, e.g. Piatti et al. (2001) list only 16 SMC clusters 
with known ages and metallicities. 
A few more have been recently added by the same authors (Piatti et al. 2005). 
Detailed CMDs for SMC clusters are still very scarce as well, which prevents 
more reliable age, metallicity and structural information from being 
derived for them.

Therefore we are studying SMC clusters by using
photometry and CMD analysis, based on a combination of CMD
modelling and statistical tools. The confirmation of some of these
clusters as intermediate or old age ones will significantly
improve the poor census in the age range corresponding
to the age gap for the SMC clusters.
Our sample data are for Lindsay~2, and Lindsay~72, for which
preliminary estimates in the literature indicate ages of
3 to 8~Gyr and $\sim$~200~Myr for Lindsay~72 (Chiosi et al. 2006),
respectively. An age indication for Lindsay~2
was obtained from low resolution integrated spectra with the 1.5m
telescopes at the ESO and LNA observed by us and based 
on integrated magnitudes and colours from Rafelski \& Zaritsky (2005).

\section{Method}
In general the SC ages are determined by using subjective (often visual) 
isochrone fits in the CMDs (Sarajedini 1998; Rich et al. 2000, 2001),
by assuming the other parameters within known ranges: metallicity,
distance modulus and reddening. The [Fe/H] values are usually determined
by spectroscopy of red giants (CaII triplet) (Grocholski et al. 2006, 
da Costa \& Hatzidimitriou 1998, Kaiser et al. 2006).

However in this work we analyse the observed CMDs by applying the method 
developed by Kerber et al. (2002, 2005, 2007). 
This method is based on statistical comparisons between the observed
CMD and a modelled CMD.
The modelling assumes that the cluster is an SSP and uses as input 
the information on metallicity, age (given by a Padova 
isochrone; Girardi et al. 2002),  intrinsic distance 
modulus ($(m-M)_0$), reddening value (E(B-V)), Mass Function slope 
($\alpha$), and fraction of unresolved binaries (f$_{bin}$), and also 
photometric uncertainties and completeness. 
In order to proceed with the statistical correction for the field star contamination
we take advantage of the outer regions of the SOI field images, not 
covered by the target SMC clusters. 
We applied the method explained in Kerber et al. (2002). 

\begin{figure}[!ht]
\centering
\includegraphics[width=0.5\textwidth]{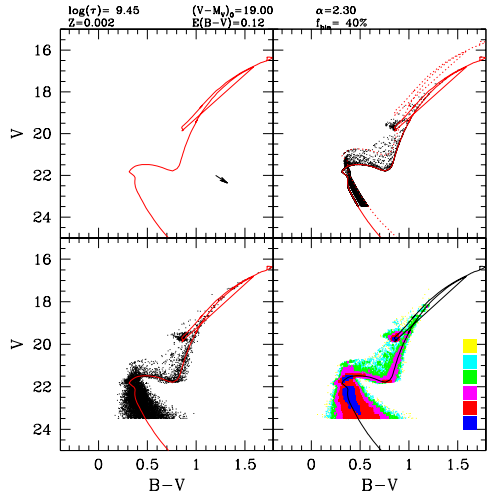}
\caption{Example of CMD modelling. {\it Panel a}: adopted isochrone (age and Z) 
shifted by distance modulus and reddening. {\it Panel b}: the distribution 
of stars in accordance to the IMF and the fraction of binaries. {\it Panel c}: 
Introduction of photometric errors and incompleteness. {\it Panel d}: stars 
codded in colour in accordance to the density of points ($\sim p_{CMD,i}$) 
in the CMD in a logarithmic scale. The adopted input physical 
parameters are indicated in the figure.}
\label{cmdmodel}
\end{figure}

The models explore a grid of expected parameters for each cluster.
So the best models are those that maximize the likelihood
(Naylor \& Jeffries 2006; Hernandez \& Valls-Gabaud 2008),
that can be defined as
$$
{\rm Likelihood } \sim \prod_{i=1}^{N_{obs}}{p_{CMD,i}} 
\sim \prod_{i=1}^{N_{obs}}{N[V_i,(B-V)_i]}
$$
where $p_{CMD,i}$ is the probability for this model to generate 
a star in the position of the i$^{th}$ observed star, and $N[V_i,(B-V)_i]$ is
the density of stars in this position, being the product 
done over all observed stars (see figure \ref{cmdmodel}).

\section{SOAR/SOI data}

Since 2006B we observe BVI images for SMC SCs 
using the SOAR Optical Imager (SOI) mounted in the 4.1m Southern 
Astrophysical Research (SOAR) Telescope, with a seeing of 
$\sim$~0.8~arcsec and a magnitude limit of V~$\sim$~23.

Our reduction procedures were based on the SOAR/IRAF packages 
and the photometry procedures were based on the DAOPHOT/IRAF 
package (Stetson 1987). We performed aperture photometry
and then we applied the point spread function (PSF) models of some bright 
and isolated field stars to all stars, in the B and V bands.
We observed some of the Sharpee et al. (2002) standards stars
in different air masses (X) and filters in
order to correctly calibrate the magnitudes to the standard 
system of magnitudes.
We achieved as dispersions from the fits:
$\sigma_V\sim0.09$ and $\sigma_{B-V}\sim0.16$.

\section{Results for Lindsay 2}

Lindsay~2 is an intermediate-age SC with age $\sim$~2.6~Gyr and 
[Fe/H]~$\sim$~-1.3 so this cluster is a new important object to trace the 
metallicity gradient in the SMC after the reactivation 
of the star formation in the last 4~Gyr.

\begin{table}[!hb]
\centering
\caption{Result parameters for Lindsay~2 from our method.}
\label{l2res}
{\scriptsize
\begin{tabular}{c|cccccc}
\hline
\noalign{\smallskip}
& Z & [Fe/H] & log($\tau$/yr) & $\tau$(Gyr) & (m-M)$_0$ & E(B-V) \\
\noalign{\smallskip}
\hline
\noalign{\smallskip}
Modelling   &  0.0010 & -1.30 &  9.42 &  2.6 & 18.99 &  0.02\\
\noalign{\smallskip}
uncertainties & +0.0010 & +0.30 & +0.07 & +0.5 & +0.10 & +0.02 \\
            & -0.0006 & -0.40 & -0.07 & -0.4 & -0.10 & -0.02 \\
\noalign{\smallskip}
\hline
\end{tabular}
}
\end{table}

\begin{figure}[!ht]
\centering
\includegraphics[width=0.5\textwidth]{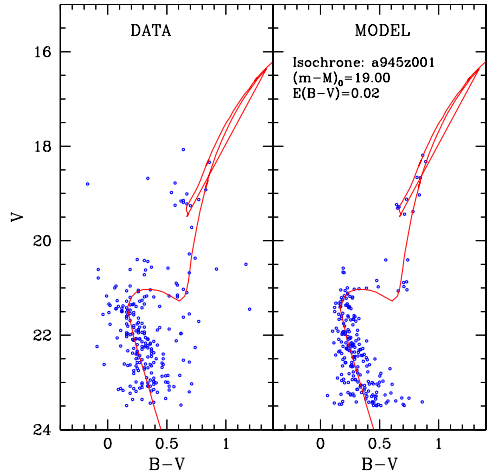}
\caption{CMDs for Lindsay~2 from observations (left) and modelling (right) with a 
Padova isochrone fit.}
\label{l2cmdmodel}
\end{figure}

\section{Conclusions \& Perspectives}
Using SOAR/SOI photometry we can objectively determine accurate and 
self-consistent physical parameters for SMC clusters by means of CMD 
modelling, as Lindsay~2 results showed.

Soon we will combine results from analysis of V,B-V and V,V-I CMDs
and we will also apply our method to Lindsay~72 and other far west SMC 
clusters that are suspect to be of intermediate or even old age.
In case we confirm that, we would be able to combine our results with those
of Crowl et al. (2001) and discuss the possibility that these 
clusters may be tidally stripped from the SMC into the Magellanic stream.


\end{document}